# Sliding Ferroelectricity Driven Spin-Layertronics in Altermagnetic Multilayers

Rui Peng[1,*], Guangxu Su[1], Yangyang Fan[1], Jiaan Li[1], Fanxin Liu[1,†], Yee Sin Ang[2,‡]

[1] School of Physics, Zhejiang University of Technology, Hangzhou 310023, China

[2] Science, Mathematics and Technology Cluster, Singapore University of Technology and Design, Singapore 487372, Singapore.

Email: [*]pengrui@zjut.edu.cn, [†]liufanxin@zjut.edu.cn, [‡]yeesin_ang@sutd.edu.sg

## Abstract

The synergy of ferroicity with altermagnetism offers a novel platform for designing multifunctional altermagnetic-spintronic device technology. In this work, we propose a mechanism to achieve nonvolatile electrical manipulation of spin and layer degrees of freedom in an altermagnetic bilayer via sliding ferroelectricity. Using first-principles calculations, we show that an interlayer translation can induce a switchable out-of-plane ferroelectric polarization in bilayer $CuF_2$, which directly couples to and reverses the *d*-wave altermagnetic spin splitting. Notably, the altermangetic spin splitting is layer-locked, the sliding ferroelectricity-driven switching thus embodying a nonvolatile spin-layertronics functionality that couples spin-polarized transport and layer degree of freedom in a single platform. We show that in quadrilayer $CuF_2$, four polarization states are identified which may offer multi-state logic device applications. These findings establish sliding ferroelectricity as a versatile tool for designing voltage-controlled, high-speed and energy-efficient spin-layertronic devices based on altermagnets.

# 1. Introduction

Altermagnetism represents an emerging frontier in collinear magnetism, uniquely combining antiferromagnetic (AFM) symmetry with a momentum-dependent spin splitting reminiscent of ferromagnets [1-5]. This novel class exhibits zero net magnetization, ensuring robustness against external fields and compatibility with high-speed operations, while simultaneously hosting a non-relativistic spin-split electronic structure that is crucial for spin manipulation [6-25]. Such a dichotomy unlocks a wealth of emergent phenomena—including the generation of non-collinear spin currents [1,6,16], anomalous Hall effects [4,10], and substantial tunneling magnetoresistance [7]—which collectively position altermagnets as a highly promising platform for next-generation spintronics. The recent experimental confirmation of various altermagnetic candidates has further amplified their potential for ultra-compact device integration [26-31]. However, a central challenge persists: achieving efficient, purely electrical control over the altermagnetic order and its associated spin polarization remains essential for their seamless integration with modern electronics.

The pursuit of electrical control in altermagnets motivates their integration with ferroic orders, leading to the emerging class of multiferroic altermagnets [32-42]. One prominent example is the altermagnetoelectric effect, where ferroelectric (FE) polarization couples directly to the altermagnetic spin texture, allowing electrical switching of the magnetic order [32-39]. This success invites the exploration of other ferroic couplings, such as with ferroelasticity, to achieve mechanical or strain-based control [40-42]. A particularly promising yet distinct avenue is the integration of altermagnetism with sliding ferroelectricy—a phenomenon unique to van der Waals multilayers, in which a lateral interlayer translation induces interfacial charge redistribution and generates a switchable out-of-plane polarization [43-52]. While major efforts have been focused on harnessing sliding ferroelectricy to manipulate unconventional magnetism in bilayer heterostructures [53-55], how the layer degree of freedom (DOF)—inherent in a multilayered system—couple cooperatively with both altermagnetic and sliding-FE orders to tunable multifunctional behavior remains unexplored thus far. Critically, the simultaneous manipulation of the layer DOF and altermagnetism via sliding ferroelectricity shall substantially broaden the physics and application potential of multiferroic altermagnets.

In this work, we propose the concept of nonvolatile switchable spin-layertronics in two-dimensional (2D) multiferroic altermagnet by integrating sliding ferroelectricity with altermagnetism and layer DOF in a bilayer setup. Using monolayer $CuF_2$—a known altermagnet exfoliable from its bulk phase [56–57]—as a building block, we show that bilayer $CuF_2$ in specific stacking configurations exhibits intrinsic sliding ferroelectricity with two stable FE states, denoted FE-I and FE-II, and sizable out-of-plane polarization ($\pm 1.23$ pC/m) that enables efficient electrical control. Notably, the system exhibit layer-locked altermagnetic spin-split bands, which are energetically shifted in the opposite fashion by the intrinsic out-of-plane FE polarization. Switching between FE-I and FE-II thus reverses the energetic shifting of these layer-locked altermagnetic spin-split bands, resulting in altermangetic spin reversal (see **Fig. 1**). By extending to quadrilayer $CuF_2$, we identify four inequivalent FE polarization states, in which the layer and the spin-polarized current can be separately controlled via sliding ferroelectricity, thus offering a spin-layertronic platform to achieve multistate device functionalities. Our findings reveal sliding-ferroelectricity-driven altermagnetic bilayer as a viable platform for electrically controllable, multi-state spin-layertronic route towards advanced memory and logic devices.

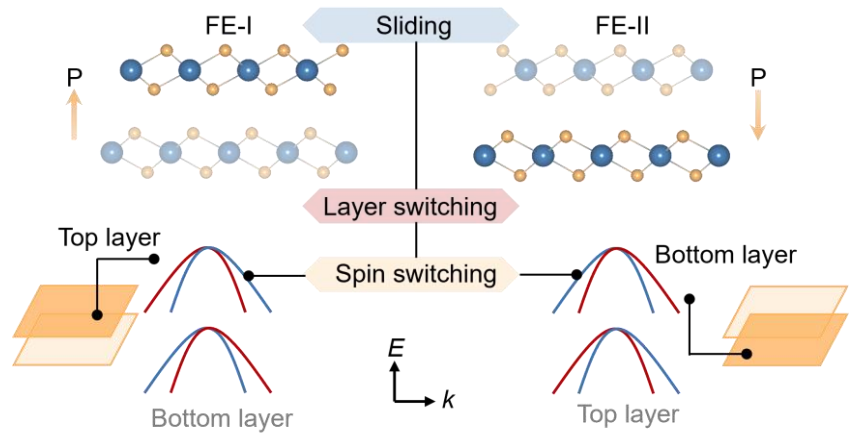

**Fig. 1** Concept of sliding-ferroelectricity driven spin-layertronic in an altermagnetic bilayer. Upon switching of the FE states (FE-I and FE-II), the built-in electric polarization is reversed. The reversal of the FE polarization causes the altermagnetic spin-split bands to energetically shift in the opposite direction for the two sub-monolayers, thus leading to altermagnetic spin switching.

## 2. Results and Discussion

### 2.1 Sliding Ferroelectricicty in Bilayer CuF$_2$

Among the commonly reported 2D altermagnetic materials, several prototypical structural motifs prevail, including the V$_2$Se$_2$O type (space group **P4/mmm**; No. 123) [6,7,12], CrS type (space group **P4/mmm**; No. 123) [8,16,17], and CuF$_2$ type (space group **P2$_1$/c**; No. 14) [9,42]; see **Fig. S1**. For the first two structural families, their stable bilayer stacking configurations crystallize in space group **P4/nmm** (No. 129), which lacks polarity and precludes the emergence of sliding ferroelectricity. In contrast, the CuF$_2$-type structure presents a distinct case. To understand the origin of the polarity in its bilayer stacking configuration, it is instructive to compare its monolayer structure with those of another two similar structures with the MX$_2$ (M = metal, X = nonmetal) stoichiometry. **Fig. S1** compares the crystal structures of monolayers 1T MoS$_2$ (space group **P$\bar{3}$m1**; No. 164), 1T' WTe$_2$ (space group **P2$_1$/m**; No. 11) and CuF$_2$. In monolayer 1T MoS$_2$, the Mo atom is octahedrally coordinated by six S atoms. In monolayer 1T' WTe$_2$, the W atoms undergo spontaneous dimerization due to Peierls distortion, resulting in a distorted octahedral coordination. In contrast, the octahedral coordination in monolayer CuF$_2$ is distorted in a different manner, driven by the Jahn–Teller effect. Recent studies demonstrated the out-of-plane switchable electric polarization in bilayer 1T' systems [44,45,47,48]. This polarization arises from uncompensated interlayer vertical charge transfer, which can be switched through interlayer sliding. Inspired by these findings, we stack monolayers CuF$_2$ into a bilayer system and explore the potential of sliding ferroelectricity in bilayer CuF$_2$.

The bilayer structure is constructed by stacking the top layer onto the bottom layer under the symmetry operation {m$_z$|(t$_x$, t$_y$, d)}, where t$_x$ and t$_y$ represent the translation distances along the x- and y-directions, respectively, and d denotes the interlayer distance. The energy landscapes of bilayer CuF$_2$ as a function of t$_x$ and t$_y$ are presented in **Fig. 2(a, b)**. As shown, the energy profile exhibits a double-well potential along the x-direction and a single-well potential along the y-direction. Energy minima occur at t$_x$ = ±a$_0$ and t$_y$ = b$_0$, where a$_0$ = 1/6a and b$_0$ = 1/2b (with a and b being the crystal constants of bilayer CuF$_2$). Correspondingly, bilayers governed by {m$_z$|(±a$_0$, b$_0$, d$_{FE}$)} correspond to two equivalent FE states, labeled FE-I and FE-II, while the configuration under {m$_z$|(0, b$_0$, d$_{IM}$)} represents an intermediate (IM)

state, with d$_{FE}$ and d$_{IM}$ referring to the interlayer distances of the FE and IM states, respectively [see **Fig. 2(c)**]. The energy barrier between the two FE states is 3.83 meV/atom. The two FE states can be reversibly switched via interlayer sliding. They can also be related to each other by a glide operation with respect to the horizontal plane.

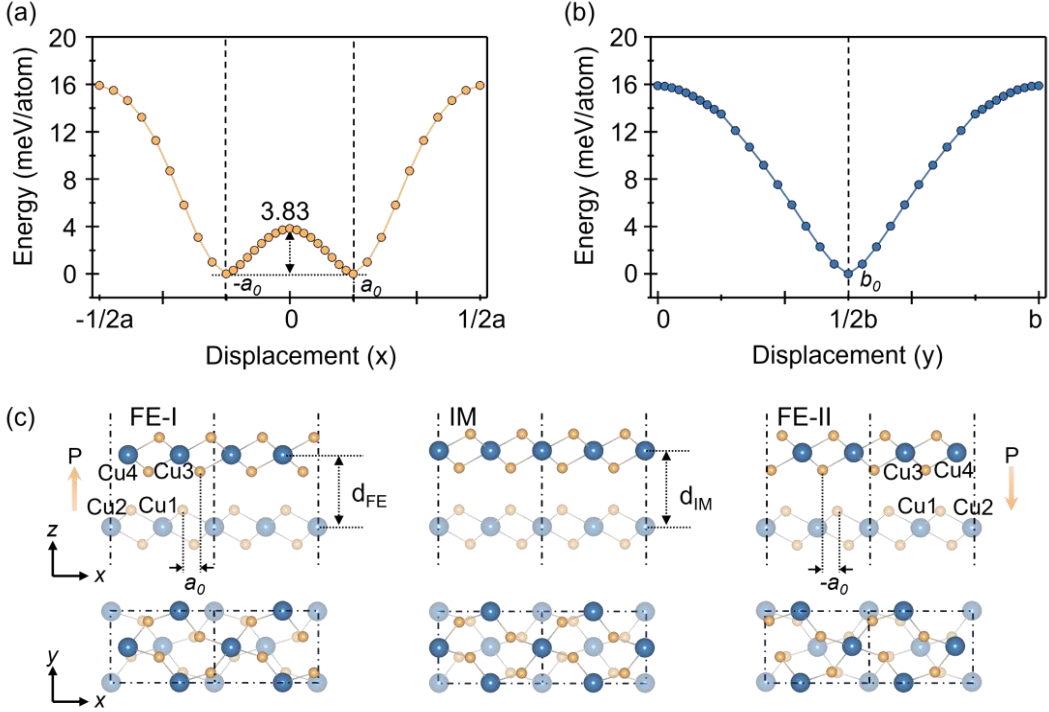

**Fig. 2** Energy profiles of bilayer CuF$_2$ along (a) x- and (b) y-directions. (c) Crystal structures of FE-I, IM, and FE-II states from side and top views.

The space group of FE-I (FE-II) state is **Pc** (No. 7), which lacks inversion symmetry. The optimized lattice constants of FE-I (FE-II) state are a = 5.32 Å and b = 3.72 Å. The asymmetry in atomic positions between the top and bottom layers results in a corresponding shift of the charge centers, which directly gives rise to the spontaneous out-of-plane electric polarization in FE-I (FE-II) state. Using the Berry phase approach, the electric polarization is calculated to be +1.23 pC/m for FE-I state and −1.23 pC/m for FE-II state (see **Fig. S2**). This magnitude exceeds that of several recently reported 2D sliding FE bilayers, such as MnBi$_2$Te$_4$ (0.45 pC/m) [49] and MX$_2$ (M = Mo, W; X = S, Se; 0.59–0.77 pC/m) [51], and is comparable to values found in B$_2$X$_3$ (X = S, Se, Te; 1.23 pC/m) [52] and BX (X = P, As, Sb; 0.965–1.075 pC/m) [50]. The presence of out-of-plane electric polarization is further corroborated by differential charge density analysis. As shown in the insets of **Fig. S2**, a clear interlayer charge transfer in FE-I state indicates an upward-oriented electric polarization. Conversely, the direction of charge transfer is reversed in FE-II state, confirming an oppositely oriented polarization. In contrast, the IM state crystallizes in the space group **Pca2$_1$** (No. 29). The specific gliding symmetry present in the IM state suppresses the out-of-plane component of the electric polarization, leading to a net zero polarization along the stacking direction.

**2.2 FE Reversal of Layer-Locked Altermagnetic Spin-Split Bands in Bilayer CuF$_2$**

In our previous work, monolayer CuF$_2$ was identified as an altermagnet [42]. To determine the magnetic ground states of bilayer CuF$_2$, we calculated the energy differences between C-type and G-type AFM orders for both the FE and IM states (see **Fig. S3**). In FE-I (FE-II) state, the C-type AFM order is found to be 0.78 meV per unit cell higher in energy than the G-type AFM order, indicating that the G-type AFM order is energetically favorable. The computed magnetic moments on the Cu sites of FE-I (FE-II) state are 0.625 (0.624), −0.625 (−0.624), −0.624 (−0.625), and 0.624 (0.625) µ$_B$ for Cu1, Cu2, Cu3, and Cu4, respectively. Due to symmetry breaking, the Cu atoms in the top (bottom) layer exhibit slightly smaller magnetic moments than those in the bottom (top) layer, although the net magnetic moment remains zero. Similarly, in the IM state, the C-type AFM order is 2.12 meV per unit cell higher than the G-type AFM, confirming the G-type order as the magnetic ground state.

In order to examine whether the synergistic switch of electric and spin/layer polarizations exist in this system, we plot the layer-resolved spin-polarized band structures and density of states for FE-I and FE-II states in **Fig. 3**, and those of IM state in **Fig. S4**. In the IM state, altermagnetic spin splitting is preserved within each layer, with opposite spin splitting between the two layers, forming a layer-locked spin-momentum coupling. The energy bands are layer-degenerate due to the absence of out-of-plane electric polarization, which leads to spin degeneracy across the entire system. From a symmetry perspective, the out-of-plane glide symmetry, when combined with the global spinor rotation {m$_z$ | 0 1/2 0}U, prohibits spin splitting. This is because the glide operation exchanges the two atoms with opposite spin—Cu1 and Cu3 (and similarly Cu2 and Cu4)—while leaving the 2D k-space invariant, thereby enforcing spin degeneracy throughout the 2D Brillouin zone.

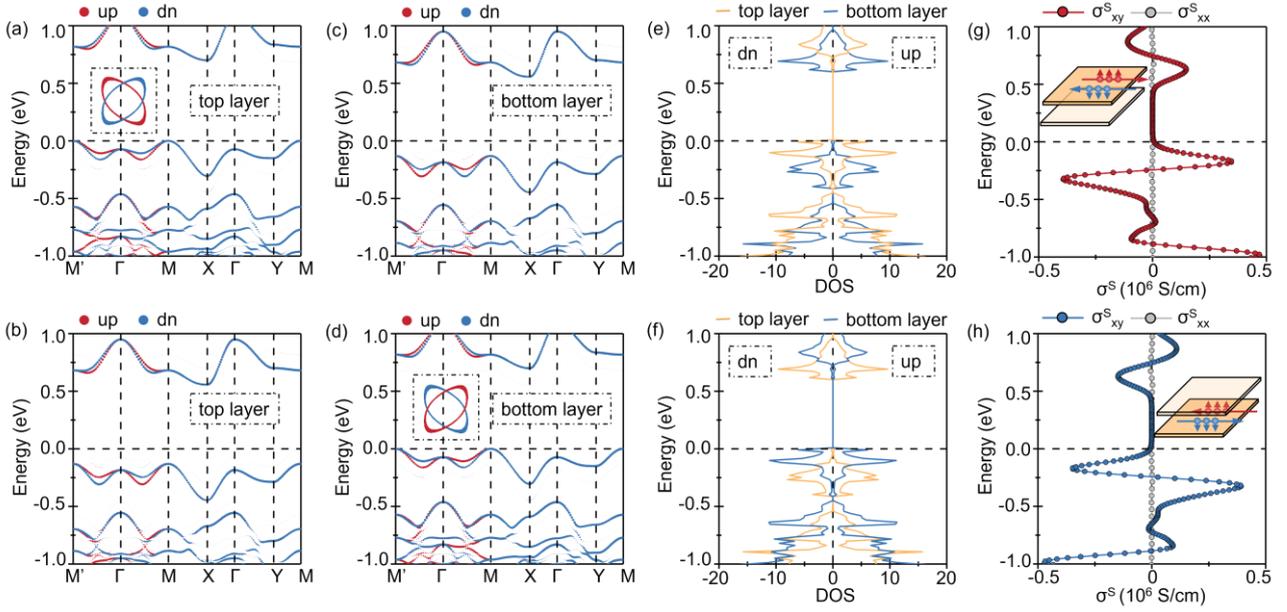

**Fig. 3** Layer-resolved spin-polarized band structures of (a) FE-I and (b) FE-II states with the contribution from the top layer. Inset in (a) is the isoenergy contours near the Fermi level of FE-I state. Layer-resolved spin-polarized band structures of (c) FE-I and (d) FE-II states with the contribution from the bottom layer. Inset in (d) is the isoenergy contours near the Fermi level of FE-II state. Density of states of (e) FE-I and (f) FE-II states. Fermi level is set to zero in (a) to (f). Longitudinal and transverse spin conductivities in (g) FE-I and (h)FE-II states. Insets are schematic diagrams of the layer-polarized spin currents in FE-I and FE-II states.

In contrast, the layer degeneracy is broken in FE-I (FE-II) state due to the presence of out-of-plane electric polarization. As shown in **Fig. 3(a-f)**, the conduction band minimum (CBM) of FE-I (FE-II) state is mainly from the bottom (top) layer, while the valence band maximum (VBM) is domanited by the top (bottom) layer. The layer polarization between the top and bottom layers is measured as 0.13 (−0.13) eV in FE-I (FE-II) state. This results in a net altermagnetic spin splitting in the band structures of both FE-I and FE-II states, resembling that of monolayer $CuF_2$. The spin-up and spin-down states remain degenerate along M–X–Γ–Y–M, while altermagnetic spin splitting emerges along M′–Γ–M. Furthermore, the spin-momentum coupling pattern is opposite between FE-I and FE-II states. Upon FE switching, the spin polarization is simultaneously reversed, enabling nonvolatile control of the altermagnetic spin splitting. This spin reversal is also evident in the isoenergy contours of FE-I and FE-II states, further confirming the feasibility of manipulating altermagnetic spin splitting through FE polarization switching, as illustrated in the insets of **Fig. 3(a) and (d)**.

**2.3 Sliding Ferroelectricity Driven Spin-Layertronics in Multilayer CuF$_2$**

The reversal of altermagnetic spin splitting leads to concomitant changes in other physical properties. In altermagnets, an anisotropic spin splitting allows an external in-plane electric field to generate a spin current [1,6,16]. The spin conductivity of bilayer CuF$_2$ can be calculated using the Boltzmann transport equation. As illustrated in **Fig. 3(g)**, the longitudinal spin conductivity in FE-I state is zero, while a sizable signal appears in the transverse direction. Notably, upon FE switching, the conductivity reverses sign owing to the opposite spin-momentum coupling patterns between FE-I and FE-II states, as shown in **Fig. 3(h)**. These results demonstrate that FE polarization can effectively control spin currents in bilayer CuF$_2$, enabling nonvolatile spin information encoding and electric-field-controllable spintronic device applications.

It is noteworthy that the layer polarization is accompanied by spin polarization, which in turn results in a layer-polarized spin current. Specifically, in the FE-I and FE-II states, the spin current is predominantly localized in the top and bottom layers, respectively, thereby endowing the spin current with an additional layer DOF; see the insets in **Fig. 3(g-h)**. Consequently, the layer-spin polarization can alternatively be detected by examining the layer localization of the spin current. We describe a state exhibiting coexisting spin and layer polarization that is switchable via FE polarization as (P, S, L), where P = $\pm1$ corresponds to upward and downward electric polarization, S = $\pm1$ represents spin-up and spin-down currents, and L = $\pm1$ denotes the top and bottom layers. Accordingly, the FE-I and FE-II states can be expressed as (+1,+1,+1) and (−1,−1,−1), respectively, implying the synergistic regulation of spin current and layer occupation by ferroelectric switching.

The interplay between electric, spin and layer polarization can be further extended to multilayer systems. For instance, in quadrilayer CuF$_2$, interlayer sliding gives rise to four inequivalent polarization states: (+2,+1,+2), (+1,−1,−1), (−1,+1,+1), and (−2,−1,−2); see **Fig. S5**. The (+2,+1,+2) state, characterized by consecutive AA stacking across two FE-I configurations, naturally exhibits upward electric polarization. Applying a glide operation to this state yields the (−2,−1,−2) state, which reverses the out-of-plane electric polarization direction. Similarly, the (+1,−1,−1) state, formed by AA stacking across FE-I and FE-II configurations, also shows upward electric polarization, and its glide counterpart (−1,+1, +1) state reverses the electric polarization. Owing to their distinct stacking sequences, the absolute electric polarization values differ between the (±2,±1,±2) and (±1,∓1,∓1) states. (±2,±1,±2) states exhibit a larger electric polarization of ±4.22 pC/m, whlie (±1,∓1,∓1) states have an electric polarization of ±1.45 pC/m. **Fig. 4** displays the band structures of these four states in quadrilayer CuF$_2$. In the (±2,±1,±2) states, the VBM originates from the two outer layers, whereas in the (±1,∓1,∓1), the VBM arises from the two inner layers. This layer-dependent band-edge localization enables selective manipulation of spin currents within specific layers, opening avenues for electric-controlled spin-layertronic functionality across all four constituent layers.

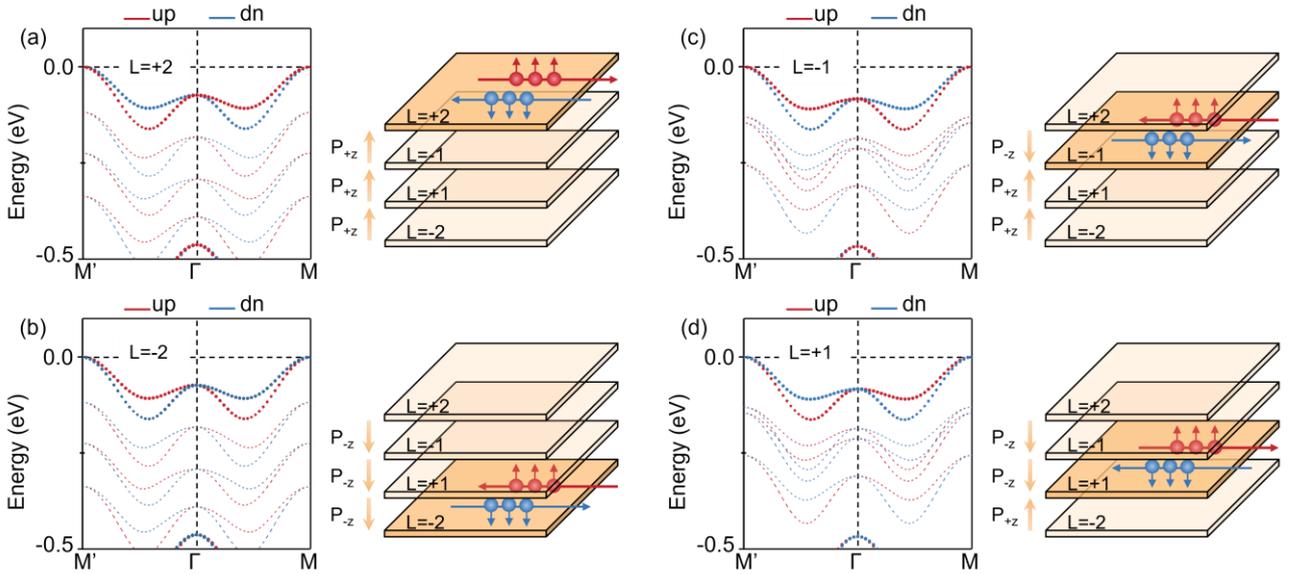

**Fig. 4** Layer-resolved spin-polarized band structures and schematic diagrams of layer-polarized spin current in (a) (+2,+1,+2), (b) (+1,−1,−1), (c) (−2,−1,−2), and (d) (−1,+1,+1) states. Fermi level is set to zero.

## 3. Conclusion

In summary, we have proposed and theoretically validated a strategy to achieve nonvolatile FE-switchable spin-layertronics by integrating altermagnetism with sliding ferroelectricity in multilayer heterostructures. We demonstrated that interlayer sliding induces out-of-plane FE polarization in bilayer CuF$_2$ which directly couples to and reverses the layer-locked d-wave altermagnetic spin-splitting texture. Transport calculations revealed that the corresponding layer-locked spin conductivities can be ferroelectrically encoded, highlighting the potential of sliding FE altermagnets as nonvolatile nanomechanical spin-layer switches. Extending to quadrilayer $CuF_2$, we identify four inequivalent polarization states, thus revealing a material platform for electrically controllable, multi-state spin-layer logic and memory devices. Our work not only addresses the key challenge of multiferroic switching in altermagnets, but also establishes FE control of spin-layertronic that enriches the device application landscape of altermangets.

## 4. Methods

First-principles calculations are performed based on density functional theory (DFT) [58] as implemented in Vienna ab initio simulation package (VASP) [59]. Exchange-correlation interaction is described by the Perdew-Burke-Ernzerhof (PBE) parametrization of generalized gradient approximation (GGA) [60]. Structures are relaxed until the force on each atom is less than 0.01 eV/Å. The cutoff energy and electronic iteration convergence criterion are set to 500 eV and $10^{-5}$ eV, respectively. To sample the 2D Brillouin zone, a Monkhorst–Pack (MP) k-grid mess [61] of 5 × 9 × 1 is used for bilayer $CuF_2$. To avoid the interaction between adjacent layers, a vacuum space of 20 Å is added. FE switching barrier is obtained by nudged elastic band (NEB) method [62]. Berry phase approach is employed to evaluate the out-of-plane electric polarization [63]. The spin-resolved transport properties are calculated using a housing-made code, in which the electron energy and electron group velocity are evaluated from the Wannier-based tight binding Hamiltonian [64].

## Supporting Information

Supporting Information is available from the Wiley Online Library or from the author.

## Acknowledgements

This work is funded by the National Natural Science Foundation of China (Nos. 12304431), and Natural Science Foundation of Zhejiang Province (Nos. LQ24A040015). Y. S. A. acknowledges the supports from the Kwan Im Thong Hood Cho Temple Early Career Chair

Professorship in Sustainability and the Singpaore Ministry of Education Academic Research Fund Tier 2 (MOE-T2EP50224-0021).

## Conflict of Interest

The authors declare no conflict of interest.

## Author Contributions

R.P. and Y.S.A. conceived and designed the project. R.P. performed the calculations and analysis. R.P. and Y.S.A. wrote the manuscript with inputs from all authors. F.X.L. reviewed and analysed the results. All the authors discussed the contents and prepared the manuscript.

## Data Availability Statement

The data that support the findings of this study are available from the corresponding author upon reasonable request.

## Keywords

Altermagnetism, Sliding ferroelectricity, Layer degree of freedom, First-principles calculations, Nonvolatile switching